\documentclass[review]{elsarticle}

\journal{Journal of \LaTeX\ Templates}

\begin{document}

\begin{frontmatter}

\title{Gapless Spin Excitations 
in the $S=1/2$ Kagome- and Triangular-Lattice 
Heisenberg 
Antiferromagnets}

\author[mymainaddress,mysecondaryaddress]{T\^oru Sakai
\corref{mycorrespondingauthor}}
\ead[url]{http://cmt.spring8.or.jp/}
\ead{sakai@spring8.or.jp}
\cortext[mycorrespondingauthor]{Corresponding author}

\author[mymainaddress]{Hiroki Nakano}



\address[mymainaddress]{Graduate School of Material Science, University of Hyogo, Hyogo 678-1297, Japan }
\address[mysecondaryaddress]{National Institutes for Quantum and Radiological Science and Technology (QST) SPring-8, Hyogo 679-5148, Japan}

\begin{abstract}
The $S=1/2$ kagome- and triangular-lattice
Heisenberg 
antiferromagnets are investigated 
using the numerical exact diagonalization and
the finite-size scaling analysis.
The behaviour of the field derivative at zero magnetization
is examined for both systems. 
The present result indicates that the spin excitation is gapless 
for each system. 
\end{abstract}


\end{frontmatter}


\section{Introduction}

Frustration in magnets is one of important topics in the field 
of the strongly correlated electron systems. 
Among such magnets, the kagome- and 
triangular-lattice antiferromagnets attract a lot of interests. 
Since discoveries of several candidate materials of the kagome-lattice 
antiferromanget; 
the herbertsmithite\cite{herbertsmithite_jacs,herbertsmithite_jpsj}, 
the volborthite\cite{volborthite_jpsj_let,volborthite_prl} 
and the vesignieite\cite{vesignieite_let}, particularly, 
the study on this system has been accelerated. 
The quantum spin-fluid behaviour of the system was predicted 
by many theoretical studies\cite{sachdev,miyashita,Lecheminant,
Waldtmann,mila,hermele,singh,ran,jiang,Cepas,Sindzingre,vidal,capponi}. 
The $U(1)$ Dirac spin-liquid theory\cite{ran} indicated 
a gapless spin excitation in the thermodynamic limit, 
which has been supported 
by the recent variational approach\cite{iqbal1,iqbal2}. 
Our recent numerical diagonalization study\cite{nakano-sakai1} also 
concluded that the system is gapless. On the other hand, 
the recent density matrix renormalization group (DMRG) 
analyses\cite{yan,depenbrock,nishimoto} suggested that 
the system has a finite spin gap even in the thermodynamic limit 
and supported the $Z_2$ topological spin-liquid picture\cite{sachdev}. 
Thus whether the $S=1/2$ kagome-lattice antiferromagnet has 
a spin gap or not is still theoretically controversial, 
although the recent neutron scattering experiment 
of the single crystal of the herbertsmithite\cite{han1,han2} suggested that 
the system is gapless. 

On the other hand, the triangular-lattice antiferromagnet 
is widely believed to be gapless,
based on the previous precise numerical analysis\cite{bernu}. 
Thus it would be interesting to compare the low-lying spin excitation
of the kagome-lattice antiferromagnet 
with the one of the triangular lattice. 
In this paper, using the recently developed field-derivative analysis 
based on the numerical diagonalization 
of finite-size clusters\cite{susceptibility}, 
we try to approach the spin-gap issue 
of the $S=1/2$ kagome-lattice antiferromagnet, 
as well as the triangular-lattice one. 
When one examines field-derivatives of the magnetization 
within the numerical data for finite-size systems, 
it is difficult to eliminate finite-size effects completely. 
It is therefore required to reduce such effects 
by means of a feasible way. 
Under these circumstances,
the purpose of this study is
to present such a field-derivative analysis
based on numerical data whose finite-size deviations are presumably reduced. 
The kagome- and triangular-lattice antiferromagnets 
have wide diversity of further studies in various aspects. 
Under large magnetic fields, nontrivial anomalous behaviours 
are observed in their magnetization curves; 
however, the behaviours are different 
between the kagome- and triangular-lattice antiferromagnets
\cite{HN_TS_kgm_ramp,nakano-sakai-kgm-39,HN_TS_kgm_1_3}. 
Such behaviors were also examined 
in various frustrated magnets\cite{HN_TS_shuriken,HN_YH_TS_spinflop,
HN_MI_TS_cairo,MI_HN_TS_cairo,MI_HN_TS_tri2dice}. 
A randomness effect in these systems 
was additionally examined\cite{watanabe_random,shimokawa_random}. 
Under such circumstances, 
the present study tackles a fundamental issue concerning 
properties of systems without effects owing to 
significantly large fields and randomness are investigated. 

\section{Model and Calculation}

Using the numerical exact diagonalization of finite-size clusters
under periodic boundary condition,
we investigate the $S=1/2$ kagome- and triangular-lattice 
Heisenberg 
antiferromagnets
defined by the Hamiltonian 
\begin{eqnarray}
\label{ham}
{\cal H} = \sum _{\langle i,j \rangle} {\bf S}_i \cdot {\bf S}_j, 
\end{eqnarray}
where site $i$ is assumed to be the vertices 
of the kagome or triangular lattice. 
Here, $\langle i,j \rangle$ runs over all the nearest-neighbor pairs 
on each lattice. 
For an $N$-site system, 
we consider subspaces characterized by $M=\sum _j S_j^z$;  
we obtain the lowest energy denoted by $E(N,M)$ 
of the Hamiltonian matrix in each subspace. 
We calculate 
all the values of $E(N,M)$ available for the 
clusters up to $N=$36 by the numerical diagonalization. 
The diagonalization is carried out based 
on the Lanczos algorithm and/or the Householder algorithm. 
Part of the Lanczos diagonalizations were carried out 
using an MPI-parallelized code which was originally developed 
in the study of Haldane gaps\cite{HaldaneGaps}. 
The usefulness of our program was confirmed 
in large-scale parallelized calculations\cite{nakano-sakai1,HN_TS_kgm_1_3,
HN_STodo_TSakai_JPSJ_S1tri,HN_TS_kgm_S,HN_YH_TS_shuriken_dist}. 

\section{Field-derivative analysis}

In order to investigate the low-lying spin excitation, 
we apply the field-derivative analysis which was developed in 
our previous work\cite{susceptibility}. 
The argument of the analysis is briefly reviewed as follows: 
the effect of the applied external magnetic field $h$ is described by 
the Zeeman energy term
\begin{eqnarray}
{\cal H}_Z=-h\sum _j S_j^z .
\label{zeeman}
\end{eqnarray}
The energy of ${\cal H}$ per site in the thermodynamic limit is defined as 
\begin{eqnarray}
{{E(N,M)}\over N} \sim \epsilon (m) \qquad (N\rightarrow \infty), 
\label{epsilon}
\end{eqnarray}
where $m=M/(SN)$ is the magnetization 
normalized by the saturated magnetization $SN$. 
If we assume $\epsilon(m)$ is an analytic function of $m$, 
the spin excitation energy would become 
\begin{eqnarray}
E(N,M+1)-E(N,M) \sim 
\frac{1}{S} \left( 
\epsilon '(m) +\frac{1}{2 } \epsilon ''(m){1\over NS} 
+ \cdots \right) . 
\label{maggap}
\end{eqnarray}
Thus, this equation gives 
the quantity corresponding to the width of the magnetization plateau 
at $m$ as follows, 
\begin{eqnarray}
(E(N,M+1)-E(N,M))-(E(N,M)-E(N,M-1)) \sim \epsilon ''(m) \frac{1}{NS^2} . 
\label{plateau}
\end{eqnarray}
Minimizing the energy of the total Hamiltonian ${\cal H}+{\cal H}_Z$, 
the ground state magnetization curve is derived by 
\begin{eqnarray}
h=\epsilon '(m)/S .
\label{mag}
\end{eqnarray}
The field derivative of the magnetization is defined as
\begin{eqnarray}
\chi \equiv {{dm}\over{dh}} = {S\over{\epsilon''(m)}} . 
\label{chi}
\end{eqnarray}
If we assume $\chi\not=0$, namely $\epsilon ''(m)$ is finite, 
the magnetization plateau at $m$ would vanish 
in the thermodynamic limit, because of (\ref{plateau}). 
Thus a necessary condition for the existence of a magnetization plateau 
at $m$ is $\chi =0$ in the thermodynamic limit. 
Now we apply this argument for the spin gap. 
We should examine the case of  $h \rightarrow 0$ corresponding to $m=0$. 
In this case, the equation (\ref{plateau}) can be rewritten as 
\begin{eqnarray}
2\Delta _N \sim \epsilon ''(0) \frac{1}{NS^2} , 
\label{spingap}
\end{eqnarray}
where $\Delta_N=E(N,1)-E(N,0)$ is the spin gap for an $N$-spin cluster. 
Thus a necessary condition of the finite spin gap would be 
$\chi=0$ at $m=0$ in the thermodynamic limit. 

In order to estimate $\chi$ at $m=0$ from discrete data 
for finite-size systems, 
it is the most simple way to use neighboring three data 
in the form $\chi_3$ defined as
\begin{eqnarray}
\chi_3^{-1} = NS[-2E(N,M)+E(N,M+1)+E(N,M-1)] .
\label{chi3}
\end{eqnarray}
Actually, eq.~(\ref{chi3}) was used 
in our previous examination\cite{susceptibility}.  
However, a significant possibility cannot be denied 
that there remain deviations from the ideal quantity 
owing to the discreteness. 
It is expected to reduce the deviations 
if one uses neighboring five data instead of the above three 
under the assumption that $\epsilon (m)$ is analytic.  
In the method employing neighboring five data, 
we use $\chi_5$ defined as
\begin{eqnarray}
\chi_5^{-1} = &NS&\left[-\frac{5}{2}E(N,M)
+\frac{4}{3}(E(N,M+1)+E(N,M-1)) \right. 
\nonumber \\
& &\left. -\frac{1}{12}\left(E(N,M+2)+E(N,M-2)\right)\right] .
\label{chi5}
\end{eqnarray}
In the thermodynamic limit, 
$\chi_3$ should agree with $\chi_5$. 
We show the estimated $\chi_3$ and $\chi_5$ for the kagome- and 
triangular-lattice antiferromagnets in the following sections.

\section{Kagome-lattice antiferromagnet}

We investigate the field derivative 
of the magnetization $\chi$ for the $S=1/2$ kagome-lattice antiferromagnet. 

\begin{figure}
\includegraphics[width=0.8\linewidth,angle=0]{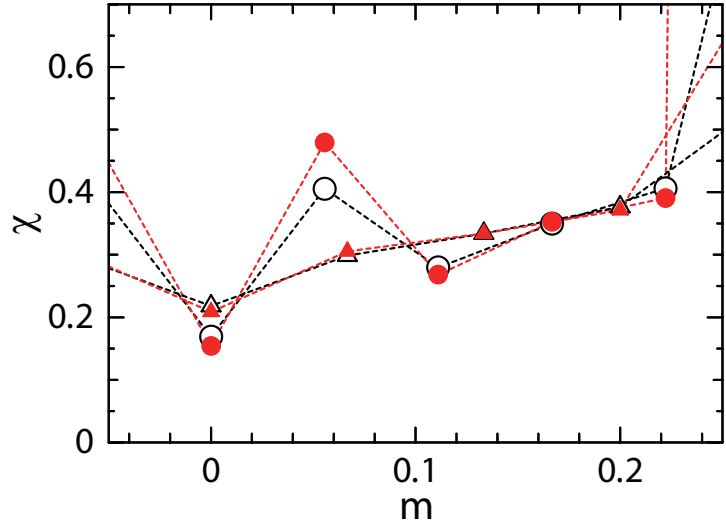}
\caption{\label{sus-m-kg} 
Field-derivative of the magnetization as a function of $m$ for $N=36$ and 30 
in the kagome-lattice antiferromagnet.
Circles and triangles denote results for $N=36$ and 30, respectively.
Black open and red closed symbols represent 
results of $\chi_3$ obtained from eq.~(\ref{chi3}) and 
$\chi_5$ obtained from eq.~(\ref{chi5}), respectively.
 }
\end{figure}

Let us, first, show differences between 
$\chi_3$ obtained from eq.~(\ref{chi3}) and 
$\chi_5$ obtained from eq.~(\ref{chi5});
results are depicted in Fig.~\ref{sus-m-kg}. 
One can confirm that the differences are small
irrespective of the values of $m$. 
The smallness is also observed irrespective of $N$. 
It is expected that 
we can obtain better estimates of $\chi$ 
by the small deviations from $\chi_3$ to $\chi_5$.  

\begin{figure}
\includegraphics[width=0.8\linewidth,angle=0]{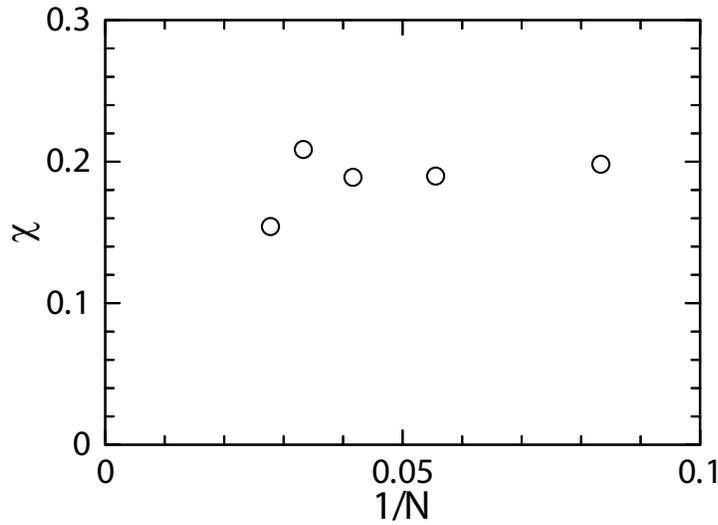}
\caption{\label{m0-kgm} 
The system-size dependence of the field derivative of the magnetization 
at $m=0$ estimated by the form (\ref{chi5}) 
in the case of the kagome-lattice antiferromagnet.  
Numerical data are plotted as a function of $1/N$
for $N$=36, 30, 24, 18, and 12.  
}
\end{figure}

Let us, next, focus our attention on the system-size dependence
of the field derivative of the magnetization 
at $m=0$. 
We plot $\chi_5$ at $m=0$ calculated by the form (\ref{chi5})
as a function of $1/N$ 
for $N$=36, 30, 24, 18, and 12 in Figure \ref{m0-kgm}. 
Although the system-size dependence exhibits a slight oscillation, 
the dependence of our data is quite small.
Our plotted data seem to go to an nonzero value 
for $N\rightarrow\infty$. 
The nonzero extrapolated value suggests that
the spin excitation of kagome-lattice antiferromagnet is gapless 
in the thermodynamic limit. 

\section{Triangular-lattice antiferromagnet}

In order to confirm the validity of the present method, 
we apply it for the $S=1/2$ triangular-lattice antiferromagnet, 
for which the triplet excitation is gapless. 

\begin{figure}
\includegraphics[width=0.8\linewidth,angle=0]{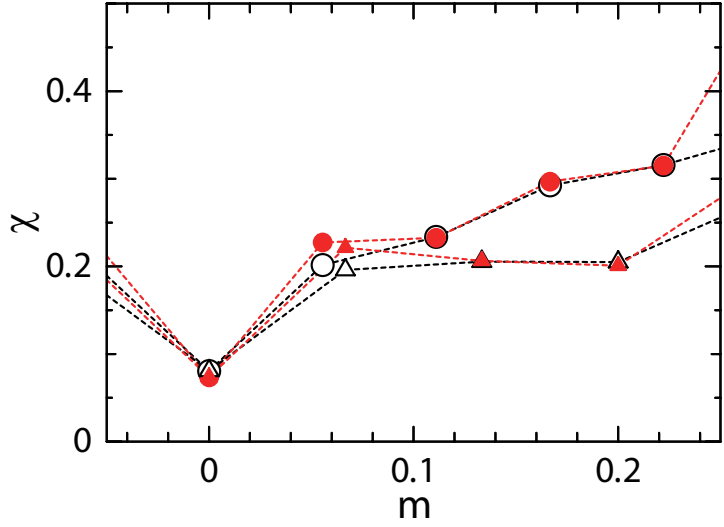}
\caption{\label{sus-m-tr} 
Field-derivative of the magnetization as a function of $m$ for $N=36$ and 30 
in the triangular-lattice antiferromagnet.
Circles and triangles denote results for $N=36$ and 30, respectively.
Black open and red closed symbols represent 
results of $\chi_3$ obtained from eq.~(\ref{chi3}) and 
$\chi_5$ obtained from eq.~(\ref{chi5}), respectively.
 }
\end{figure}

Figure~\ref{sus-m-tr} presents differences between 
$\chi_3$ obtained from eq.~(\ref{chi3}) and 
$\chi_5$ obtained from eq.~(\ref{chi5});
results are given for $N=36$ and 30. 
One can confirm that the differences are small
irrespective of the values of $m$. 
The smallness is also observed irrespective of $N$. 
Note here that the differences are even smaller than in the case
of the kagome-lattice antiferromagnet in Fig.~\ref{sus-m-kg}. 

\begin{figure}
\includegraphics[width=0.8\linewidth,angle=0]{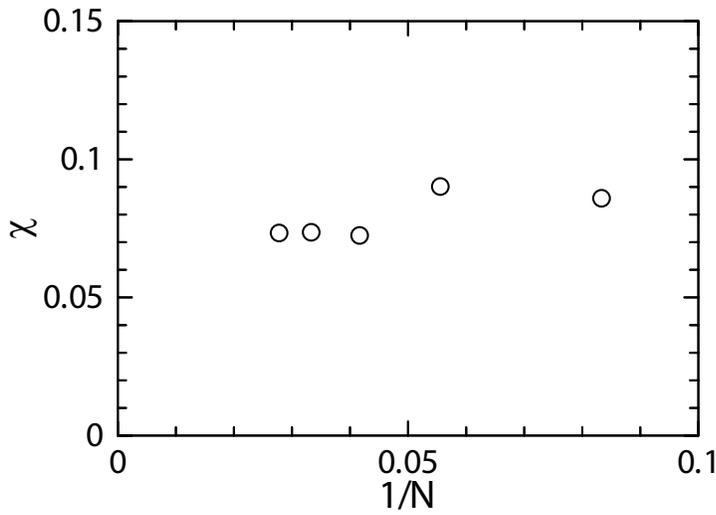}
\caption{\label{m0-tri} 
The system-size dependence of the field derivative of the magnetization 
at $m=0$ estimated by the form (\ref{chi5}) 
in the case of the triangular-lattice antiferromagnet.  
Numerical data are plotted as a function of $1/N$
for $N$=36, 30, 24, 18, and 12.  
 }
\end{figure}

Figure~\ref{m0-tri} provides us with the system-size dependence
of the field derivative of the magnetization at $m=0$;
$\chi_5$ at $m=0$ calculated by the form (\ref{chi5}) are plotted 
as a function of $1/N$ for $N$=36, 30, 24, 18, and 12. 
The oscillating behaviour in the system-size dependence
seems smaller than in the case of the kagome-lattice antiferromagnet. 
The result of the dependence indicates 
that the system has a non-zero field derivative at $m=0$
in the thermodynamic limit. 
The nonzero extrapolated value suggests that
the spin excitation of triangular-lattice antiferromagnet is gapless, 
which is consistent with a widely believed consensus of the gapless feature 
of the triangular-lattice antiferromagnet. 
Thus it is confirmed that the present method is valid.

\section{Summary}

The low-lying spin excitations of 
the kagome- and triangular-lattice antiferromagnets are investigated 
using the numerical diagonalization up to $N=36$. 
The analysis of the field derivative of the magnetization $\chi$ 
is successfully applied to these two systems and we conclude 
that these systems are gapless in the spin excitation. 
Our numerical results strongly suggest that the gapless spin excitation 
in the kagome-lattice antiferromagnet, as well as the triangular-lattice one. 
The study presents an analysis based on numerical data 
whose finite-size effects are expected to be reduced 
because $\chi$ is estimated from the neighboring five data 
of the ground-state energies. 
The present method would be applicable for various frustrated quantum spin 
systems. 

\section*{Acknowledgments}

This work has been partly supported by Grants-in-Aids for Scientific Research 
(Nos. 16H01080(JPhysics), 16K05418, and 16K05419) from the Ministry of Education, Culture, Sports, Science and Technology of Japan (MEXT), and Hyogo Science and Technology Association. This research used computational resources of the K computer provided by the RIKEN Advanced Institute for Computational Science through the HPCI System Research projects (Project IDs: hp130070, hp130098, hp150024, hp170017, hp170028, hp170070 and hp170207). We further thank the Supercomputer Center, Institute for Solid State Physics, University of Tokyo; the Cyberscience Center, Tohoku University; and the Computer Room, Yukawa Institute for Theoretical Physics, Kyoto University; the Department of Simulation Science, National Institute for Fusion Science; Center for Computational Materials Science, Institute for Materials Research, Tohoku University; Supercomputing Division, Information Technology Center, The University of Tokyo for computational facilities. This work was partly supported by the Strategic Programs for Innovative Research, MEXT, and the Computational Materials Science Initiative, Japan. The authors would like to express their sincere thanks to the staff members of the Center for Computational Materials Science of the Institute for Materials Research, Tohoku University for their continuous support of the SR16000 supercomputing facilities.




\begin{thebibliography}{9}

\bibitem{herbertsmithite_jacs}
M.~P.~Shores, E.~A.~Nytko, B.~M.~Barlett, and D.~G.~Nocera, 
J.~Am.~Chem.~Soc. {\bf 127}, 13462 (2005). 

\bibitem{herbertsmithite_jpsj}
P.~Mendels and F.~Bert, 
J.~Phys.~Soc.~Jpn. {\bf 79}, 011001 (2010).

\bibitem{volborthite_jpsj_let} 
H.~Yoshida, Y.~Okamoto, T.~Tayama, 
T.Sakakibara, M.Tokunaga, A.~Matsuo, Y.~Narumi, K.~Kindo,
M.~Yoshida, M.~Takigawa, and Z.~Hiroi, 
J.~Phys.~Soc.~Jpn. \textbf{78}, 043704 (2009).
\label{volborthite_jpsj_let}

\bibitem{volborthite_prl} 
M.~Yoshida, M.~Takigawa, H.~Yoshida, Y.~Okamoto, and Z.~Hiroi, 
Phys.~Phys.~Lett. \textbf{103}, 077207 (2009).
\label{volborthite_prl}

\bibitem{vesignieite_let}
Y.~Okamoto, H.~Yoshida, and Z.~Hiroi, 
J.~Phys.~Soc.~Jpn. \textbf{78}, 033701 (2009).


\bibitem{sachdev}
S. Sachdev, Phys. Rev. B {\bf 45}, 12377 (1992). 

\bibitem{miyashita} 
T.~Nakamura and S.~Miyashita, 
Phys.~Rev.~B \textbf{52}, 9174 (1995).

\bibitem{Lecheminant} 
P.~Lecheminant, B.~Bernu, C.~Lhuillier, L.~Pierre, and 
P.~Sindzingre, 
Phys.~Rev.~B \textbf{56}, 2521 (1997).

\bibitem{Waldtmann} 
Ch.~Waldtmann, H.-U.~Everts, B.~Bernu, C.~Lhuillier, 
P.~Sindzingre, P.~Lecheminant and L.~Pierre, Eur.~Phys.~J.~B 
\textbf{2}, 501 (1998).

\bibitem{mila}
F. Mila, Phys. Rev. Lett. {\bf 81}, 2356 (1998).

\bibitem{hermele}
M. Hermele, T. Senthil and M. P. A. Fisher, Phys. Rev. B {\bf 72}, 
104404 (2005). 

\bibitem{singh}
R. R. P. Singh and D. A. Huse, Phys. Rev. B {\bf 76}, 
180407(R) (2007).

\bibitem{ran}
Y. Ran, M. Hermele, P. A. Lee and X. -G. Wen, Phys. Rev. Lett. 
{\bf 98}, 117205 (2007).

\bibitem{jiang}
H. C. Jiang, Z. Y. Weng and D. N. Sheng, Phys. Rev. Lett. 
{\bf 98}, 117203 (2008).



\bibitem{Cepas}
O.~Cepas, C.~M.~Fong, P.~W.~Leung, and C.~Lhuillier,  
Phys.~Rev.~B \textbf{78}, 140405(R) (2008).

\bibitem{Sindzingre}
P.~Sindzingre and C.~Lhuillier, 
Europhys.~Lett. \textbf{88}, 27009 (2009). 

\bibitem{vidal}
G. Evenbly and G. Vidal, Phys. Rev. Lett. {\bf 104}, 187203 (2010).

\bibitem{capponi}
S. Capponi, O. Derzhko, A. Honecker, A. M. L\"auchli and J. Richter, Phys. Rev. B {\bf 88}, 144416 (2013).

\bibitem{iqbal1}
Y. Iqbal, D. Poilblanc and F. Becca, Phys. Rev. B {\bf 89}, 020407(R) (2014).

\bibitem{iqbal2}
Y. Iqbal, D. Poilblanc and F. Becca, Phys. Rev. B {\bf 91}, 020402(R) (2015).

\bibitem{nakano-sakai1}
H. Nakano and T. Sakai, J. Phys. Soc. Jpn. {\bf 80}, 053704 (2011). 

\bibitem{yan}
S. Yan, D. A. Huse and S. R. White, Science {\bf 332}, 1173 (2011).

\bibitem{depenbrock}
S. Depenbrock, I. P. McCulloch and U. Schollw\"ock, Phys. Rev. Lett. 
{\bf 109}, 067201 (2012).

\bibitem{nishimoto}
S. Nishimoto, N. Shibata and C. Hotta, Nat. Commun. {\bf 4}, 2287 (2013).



\bibitem{han1}
T. Han {\it et al}., Nature Commun. {\bf 492}, 406 (2012).

\bibitem{han2}
T. Han {\it et al}., Phys. Rev. Lett. {\bf 108}, 157202 (2012).

\bibitem{bernu}
B. Bernu, P. Lecheminant, C. Lhuillier and L. Pierre, 
Phys. Rev. B {\bf 50}, 10048 (1994). 

\bibitem{susceptibility}
T. Sakai and H. Nakano, Polyhedron {\bf 126}, 42 (2017).

\bibitem{HN_TS_kgm_ramp}
H.~Nakano and T.~Sakai, 
J. Phys. Soc. Jpn. \textbf{79}, 053707 (2010).

\bibitem{nakano-sakai-kgm-39}
T. Sakai and H. Nakano, 
Phys. Rev. B {\bf 83}, 100405(R) (2011).

\bibitem{HN_TS_kgm_1_3}
H.~Nakano and T.~Sakai, 
J. Phys. Soc. Jpn. \textbf{83}, 104710 (2014).

\bibitem{HN_TS_shuriken}
H.~Nakano and T.~Sakai, 
J. Phys. Soc. Jpn. \textbf{82}, 083709 (2013).

\bibitem{HN_YH_TS_spinflop}
H.~Nakano, Y.~Hasegawa, and T.~Sakai, 
J. Phys. Soc. Jpn. \textbf{83}, 084709 (2014).


\bibitem{HN_MI_TS_cairo}
H.~Nakano, M.~Isoda, and T.~Sakai, 
J. Phys. Soc. Jpn. \textbf{83}, 053702 (2014).

\bibitem{MI_HN_TS_cairo}
M.~Isoda, H.~Nakano, and T.~Sakai, 
J. Phys. Soc. Jpn. \textbf{83}, 084710 (2014).

\bibitem{MI_HN_TS_tri2dice}
H.~Nakano and T.~Sakai, 
J. Phys. Soc. Jpn. \textbf{86}, 063702 (2017).

\bibitem{watanabe_random}
K.~Watanabe, H.~Kawamura, H.~Nakano, and T.~Sakai, 
J. Phys. Soc. Jpn. \textbf{83}, 034714 (2014)

\bibitem{shimokawa_random}
T.~Shimokawa, K.~Watanabe, and H.~Kawamura, 
Phys. Rev. B \textbf{92}, 134407 (2015). 

\bibitem{HaldaneGaps}
H.~Nakano and A.~Terai, 
J. Phys. Soc. Jpn. \textbf{78}, 014003 (2009).

\bibitem{HN_STodo_TSakai_JPSJ_S1tri}
H.~Nakano, S.~Todo, and T.~Sakai, 
J. Phys. Soc. Jpn. \textbf{82}, 043715 (2013).

\bibitem{HN_TS_kgm_S}
H.~Nakano and T.~Sakai, 
J. Phys. Soc. Jpn. \textbf{84}, 063705 (2015).

\bibitem{HN_YH_TS_shuriken_dist}
H.~Nakano, Y.~Hasegawa, and T.~Sakai, 
J. Phys. Soc. Jpn. \textbf{84}, 114703 (2015).

\end{thebibliography}

\end{document}